\documentclass[manuscript,screen]{acmart}
\AtBeginDocument{%
  \providecommand\BibTeX{{%
    \normalfont B\kern-0.5em{\scshape i\kern-0.25em b}\kern-0.8em\TeX}}}

\setcopyright{acmlicensed}
\copyrightyear{2024}
\acmYear{2024}
\acmDOI{XXXXXXX.XXXXXXX}

\acmConference[CHI '24]{May 12, 2024}{Honolulu, HI}
\acmBooktitle{CHI '24: With or Without Permission,
 May 12, 2024, Honolulu, HI} 
\acmISBN{978-1-4503-XXXX-X/18/06}

\begin{document}

\title{Sociotechnical Considerations for SLAM Anchors in Location-Based AR}

\author{Tiffany T. Nguyen}
\email{tnguyen26@scu.edu}
\orcid{0009-0009-7727-2796}
\affiliation{%
  \institution{Santa Clara University}
  \streetaddress{500 El Camino Real}
  \country{USA}
  \postcode{95053}
}

\author{Cinthya Jauregui}
\email{cjauregui@scu.edu}
\orcid{0009-0008-2860-3264}
\affiliation{%
  \institution{Santa Clara University}
  \country{USA}
}

\author{Sarah H. Sallee}
\email{ssallee@scu.edu}
\orcid{0009-0006-5241-0791}
\affiliation{%
  \institution{Santa Clara University}
  \streetaddress{500 El Camino Real}
  \country{USA}
  \postcode{95053}
}

\author{Mohan R. Chandrasekar}
\email{mchandrasekar@scu.edu}
\orcid{0009-0000-3860-4609}
\affiliation{%
  \institution{Santa Clara University}
  \streetaddress{500 El Camino Real}
  \country{USA}
  \postcode{95053}
}

\author{Liam A'Hearn}
\email{lahearn@scu.edu}
\orcid{0009-0005-9119-9186}
\affiliation{%
  \institution{Santa Clara University}
  \streetaddress{500 El Camino Real}
  \country{USA}
  \postcode{95053}
}

\author{Dominic J. Woetzel}
\email{dwoetzel@scu.edu}
\orcid{0009-0006-2956-6237}
\affiliation{%
  \institution{Santa Clara University}
  \streetaddress{500 El Camino Real}
  \country{USA}
  \postcode{95053}
}

\author{Pinak Paliwal}
\email{pinakpaliwal@gmail.com}
\orcid{0009-0002-6623-4849}
\affiliation{%
  \institution{Santa Clara University}
  \streetaddress{500 El Camino Real}
  \country{USA}
  \postcode{95053}
}

\author{Madison Nguyen}
\email{mnguyen9@scu.edu}
\orcid{0009-0003-2779-5050}
\affiliation{%
  \institution{Santa Clara University}
  \streetaddress{500 El Camino Real}
  \country{USA}
  \postcode{95053}
}

\author{Isabella `Amne Gomez}
\email{iagomez@scu.edu}
\orcid{0009-0003-2289-4279}
\affiliation{%
  \institution{Santa Clara University}
  \streetaddress{500 El Camino Real}
  \country{USA}
  \postcode{95053}
}

\author{Xinqi Zhang}
\email{xzhang22@scu.edu}
\orcid{0009-0004-7493-9800}
\affiliation{%
  \institution{Santa Clara University}
  \streetaddress{500 El Camino Real}
  \country{USA}
  \postcode{95053}
}

\author{Lee M. Panich}
\email{lpanich@scu.edu}
\orcid{0000-0002-5741-8921}
\affiliation{%
  \institution{Santa Clara University}
  \streetaddress{500 El Camino Real}
  \country{USA}
  \postcode{95053}
}

\author{Danielle M. Heitmuller}
\email{dheitmuller@scu.edu}
\orcid{0009-0000-2193-4281}
\affiliation{%
  \institution{Santa Clara University}
  \streetaddress{500 El Camino Real}
  \country{USA}
  \postcode{95053}
}

\author{Amy Lueck}
\orcid{0000-0002-0432-2280}
\email{alueck@scu.edu}
\affiliation{%
  \institution{Santa Clara University}
  \country{USA}
}

\author{Kai Lukoff}
\email{klukoff@scu.edu}
\orcid{0000-0001-5069-6817}
\affiliation{%
  \institution{Santa Clara University}
  \streetaddress{500 El Camino Real}
  \country{USA}
  \postcode{95053}
}

\renewcommand{\shortauthors}{Nguyen, et al.}

\begin{abstract}
  In this position paper, we explore the power of storytelling and its connection to place through the use of Augmented Reality (AR) technology, particularly within the context of Thámien Ohlone history on the Santa Clara University campus. To do this, we utilized SLAM and 8th Wall to create virtual, location-based experiences that geolocate tribal stories at present-day sites, showcase the living culture of the Thámien Ohlone tribe, and advocate for physical markers that could exist to recognize their story. When doing so, we made sure to select locations that added to the story each stop tells to serve as our anchors. Our research then investigates both the social and technical considerations involved in selecting anchors for AR experiences, using the Thámien Ohlone AR Tour as a case study.
\end{abstract}




\begin{teaserfigure}
\begin{tabular}{c  c  c}
    \centering
     \begin{minipage}[t]{.33\textwidth}
        \centering
        \includegraphics[width=.7\linewidth]{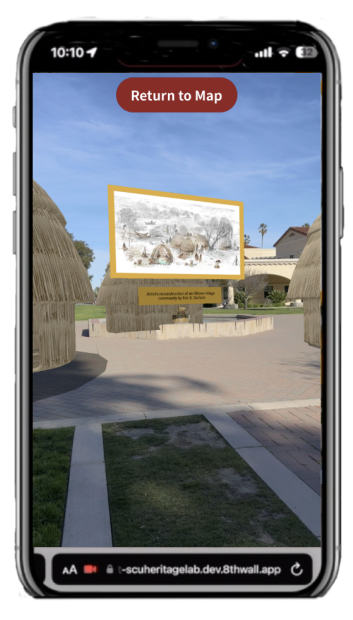}
        \caption{An AR stop showcasing Thámien Ohlone village life located around a fountain at the center of where a village once stood.}
        \label{fig:fountain}
    \end{minipage} &  
    \begin{minipage}[t]{.33\textwidth}
        \centering
        \includegraphics[width=.7\linewidth]{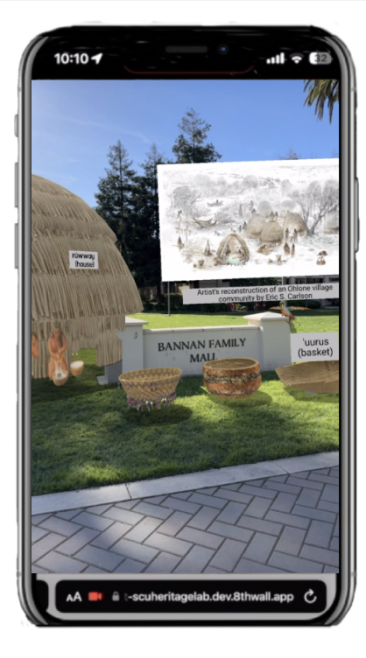}
        \caption{The Bannan Family Mall Sign acts as another potential anchor for the village stop. While it holds the same historical significance of being physically near the center of where a village once stood, the perceived message that comes from scanning a sign thanking contemporary donors can detract from the story being told.}
        \label{fig:bannan}
    \end{minipage} &
    \begin{minipage}[t]{.33\textwidth}
        \centering
        \includegraphics[width=.7\linewidth]{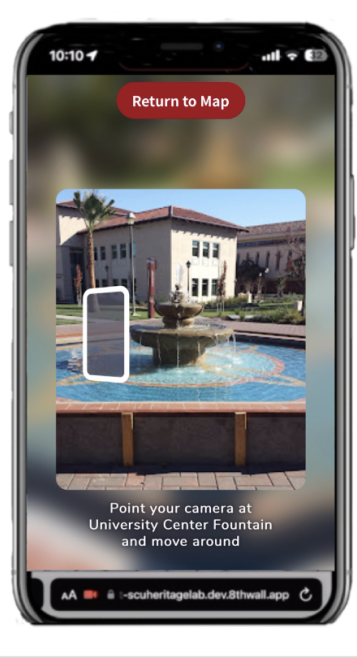}
        \caption{The tour's Web-based AR experiences require users to localize to each stop by scanning the anchor. This process actively engages the user with the stop’s anchor, so intentionally choosing anchors with a purpose can add to the stop's story.}
        \label{fig:scanning}
    \end{minipage}
\end{tabular}
\end{teaserfigure}


\maketitle

\section{Introduction}
Stories are powerful. They allow us to learn from history, connect with people across space and time, and create meaning out of our experiences. Many of the times, stories are also intrinsically tied to the locations where they originally unfolded and are recounted. 

With this perspective in mind, we aimed to create a location-based, augmented reality tour of the Santa Clara University campus highlighting Thámien Ohlone history. The Ohlone people have stewarded the land our university is built on for thousands of years, and Mission Santa Clara itself was a place native people built, lived, labored, and died in. Furthermore, the Ohlone people continue to live here today, and yet a 2020 working group report \cite{Baines2020-oj} by scholars, university administrators, and Ohlone leaders highlights there are minimal physical markers on our university’s campus highlighting the Ohlone story. To provide a counternarrative, we decided to utilize Augmented Reality to envision a more just future by showing what physical markers could exist, creating informative stops that can reach more people \cite{Schroeder2023-qn}, connecting with Indigenous traditions of place-based storytelling, and working with rapidly improving technology \cite{Silva2024-jr}. “Why here?” is a guiding question for the Thámien Ohlone AR Tour: each of the three stops on the walking tour is selected for its connection to its location \cite{Jauregui2024-ai}. A preview video of the experience is available \href{https://www.youtube.com/watch?v=i7Lmqqq2pBo}{here}.

To create this tour, we used simultaneous localization and mapping (SLAM) integrated into 8th Wall Web-AR technology. SLAM is a popular technique for enabling augmented reality (AR) experiences to appear at specific physical locations. In SLAM, objects anchor and allow centimeter-level precision. By comparison, other techniques, such as GPS-based localization, have the advantage of not requiring an anchor to be scanned, but are usually only accurate to within 5-10 feet. When AR content needs to be localized with a high degree of precision, SLAM-based localization is currently the preferred choice. 

With SLAM, users navigate to each stop’s site, where they will be shown an image of the anchor along with instructions to scan by moving their camera back and forth around the object (see Figure \ref{fig:scanning}). Once localization is successful, the website will use the user’s position relative to the anchor to display virtual objects and content, creating an immersive environment that connects the virtual world to the real one. Because the process of scanning actively engages the user with the stop’s anchor, our research asks: what are the sociotechnical considerations for choosing SLAM anchors, using the Thámien Ohlone AR Tour as a test case? At this CHI24 workshop on site-specific augmented reality for social justice \cite{Silva2024-jr}, we hope to discuss and receive feedback on the considerations we have identified so far.

\section{Social Considerations}
To fully take advantage of this localization technology, it was important for us to utilize archeological evidence to be intentional about choosing the locations that uplift the invisible stories of past tribes. Because of this, choosing anchors meant picking places that enhance the story and the experience, not just spots where the technology works well.

\subsection{Historical Significance}

The first factor we considered was the historical context behind each stop’s location and how it adds to the story. For example, we chose to anchor the village stop around a fountain on campus as pictured in Figure \ref{fig:fountain} because archaeological evidence suggests that the fountain stands at the center of where a village once stood. For us, it was especially important to highlight what is part of history and what has been erased, making the “invisible history” visible.

With this comes the power of highlighting the past, present, and future all in one location. For example, we developed a Names in the Sky stop where we displayed the names of Ohlone ancestors buried around Mission Santa Clara. Not only does this stop honor the lives of these individuals and showcase how the mission was a primarily Native site, but it also questions the lack of historical recognition through the absence of a physical marker denoting the location as a cemetery. This then pushed us to consider creating another stop focused on what markers could exist in front of the cemetery.

\subsection{Perceived Significance}
In addition to the historical implications of an anchor, it is also important to consider how the user would interact with the stop - namely, what is apparent at the present day site and what is the perceived significance of that anchor - since the historical meaning may not be apparent at first glance. 

For instance, we initially chose to anchor a stop highlighting village life to a fountain at the location where an Ohlone village once stood. When we started to face difficulties scanning the fountain because of the water’s reflective surface, we considered localizing instead to the nearby Bannan Family Mall sign that, like the fountain, is physically close to the village's original site. However, as shown in Figure \ref{fig:bannan}, we quickly realized that doing so would conflate the apparent message on the sign – contemporary donations to the university – with the purpose of the stop – to showcase pre-colonial village life. As a result, we went back to using the fountain as an anchor despite the technical challenges it posed. 

Some additional examples of potential locations for other stops we considered because of the site’s perceived significance include photos of university presidents to contrast with the role of present day tribal leaders, flagpoles on campus to emphasize the lack of recognition in how the Ohlone flag isn’t flown, the memorial site where thousands of Ohlone people were buried, and the Mission Church to reclaim the site as a Native site in addition to being a Catholic one.

\subsection{Stop Availability}
Another important social consideration to examine is the availability of the anchor as a location once the tour becomes public. This includes factors like public access, shelter from the sun/rain, how crowded the location will be, and how safe a location is. For example, we considered anchoring a stop to a parking lot with the outline of an adobe structure that was once a Native Rancheria. However, we quickly realized that using an AR tour where people will be looking at their phones may be dangerous in a parking lot with cars driving through.

When creating larger tours with multiple stops, it is also important to consider the walkability of the tour in terms of the distance between anchors and each stop’s longevity with whether the anchor will still be there after 2, 4, or 10 years.

\subsection{Privacy Concerns}
Another social consideration is privacy concerns. This was especially prevalent for us since we used Niantic’s geospatial browser, where all meshes of public wayspots are accessible for anyone to view. As a result, it is important to ensure that we are not accidentally gathering people’s personal information when scanning the stops, especially if AR spots are close to where people normally walk through or work.

\section{Technical Considerations}
\subsection{User Experience}
When designing stops with user experience in mind, it is important to consider the anchor itself since factors like symmetry can make it hard to situate the user in a specific view. For example, we anchored the village stop around a fountain with the intention of creating an open world feeling with huts around the fountain. However, the symmetry of the fountain meant users would scan different sides of the fountain and occasionally begin the experience within a hut or to the side of the image. In another instance, we had to consider if the user should be allowed to scan multiple sides of an anchor and what the experience would be like from each angle. In the Pow Wow stop, we wanted to highlight cultural revival using images scattered around a lawn where an annual Pow Wow takes place. However, the arch we were localizing to had a sidewalk on its other side, meaning that users’ inclination to walk along the sidewalk and scan that side meant that they would not be able to clearly see images on the lawn. As a result, we had to find ways to make scanning instructions more clear.

\subsection{Technical Viability}
Beyond social and UX considerations, it is also important to consider the technical viability of an anchor. The official 8th Wall guidelines discourages scanning reflective or transparent surfaces, and doing so led us to run into issues with scanning the fountain because of its flowing water. However, we were able to circumvent this by gathering scans while the fountain was off and the water was drained, allowing us to get more accurate scans of the fountain’s tiles.

When trying out different potential stops, we also found that the scanning technology had a hard time keeping track of thinner poles like fences, so we focused on choosing larger and more distinct objects to act as our anchor.

Another important consideration is how the lighting at a stop changes based on weather conditions and time of day. It is important to decide what times the tour supports, such as whether or not nighttime viewing will be available, and gathering sufficient scans at the appropriate times. 

Finally, a fourth technical consideration to keep in mind is wifi and cellular data constraints. For example, the outdoor wifi on campus was slow at some of the locations we considered, resulting in longer load times or requiring that the user use cellular data to be able to load all the 3D models.

\subsection{Privileging the Built Over the Natural Environment}
Depending on the stop’s theme, finding culturally appropriate anchors can also be challenging due to limitations in the technology, which favors distinct, built, and maintained architecture. The localization technology utilizes scans of a location captured by the developer to anchor virtual content around a physical space, and these scans are most representative of a site when the setting does not change over time. As a result, natural landscapes like plants that change with the seasons and grow across years cannot easily be used as anchors. Furthermore, users need to be able to easily navigate to and scan a specific object, so environmental features like trees that can be mistaken for one another would not work well. Because of this, localization technology disadvantages telling stories about natural landscapes, which creates limitations since many Indigenous stories are tied to natural features of the land around them, such as their relationship to specific plants or water.

\section{Co-designing with the Muwekma Ohlone Tribe}
This tour is co-designed with Ohlone people, specifically members of the Muwekma Ohlone Tribe. Our collaborative work ensures that the development and narrative of the tour are in harmony with the tribe's vision, respecting their preferences regarding the dissemination of their cultural narratives. This collaboration involved regular communication through monthly emails, bimonthly video conferences, and semiannual in-person workshops with tribal members. Moreover, we have initiated bi-weekly co-design meetings with tribal youth, aiming to integrate their perspectives and innovate our methods, particularly in incorporating physical locations into the AR experience due to the unique challenges of linking digital content with specific geographical landmarks.

\subsection{Innovating New Methods in Site-Specific Design}
One of our key methods, termed Landmark-based Affinity Diagramming, was developed to ideate and organize potential AR tour stops. This process began with identifying themes from prior discussions with the tribe to avoid redundancy, encouraging the tribe to introduce new narratives they wished to share. These themes were then physically represented with pink post-it notes and placed onto a large-scale printed map of Santa Clara University's campus. This mapping was guided by the significance of locations—accounting for both social and technical considerations as discussed earlier. Available or obtainable digital assets like 3D models, audio clips, and images were written on blue post-it notes, aligning them with the thematic stops. The prioritization of these potential stops was indicated by additional stickers, allowing us to discern the tribe's preferences for which stories to develop first.

\subsection{Collaborating with Tribal Youth}
Extending our collaborative efforts, we also focus on involving tribal youth, recognizing their unique insights into their culture. A case in point involves a 12-year-old tribal member with interests in design and technology. Our approach employs the educational strategy of "I do, we do, you do," designed to scaffold her learning experience with AR technologies. This included hands-on demonstrations of AR content creation, such as the digitization of cultural artifacts into 3D models and their spatial placement within the AR environment using the editor for the 8th Wall WebAR service. Through active engagement in the design process, she gains practical skills and a deeper understanding of how her culture can be represented through digital technology.

\section{Conclusion}
The collaborative design process for the AR tour of Santa Clara University's campus has highlighted numerous sociotechnical challenges and opportunities. By employing technologies like SLAM, we've endeavored to an Ohlone vision of the past, present, and future of their culture. Our project illustrates the potential of AR to elevate indigenous voices and contribute to a broader understanding of cultural heritage and social justice. As we continue to refine our collaborative methodologies and engage with tribal communities, we anticipate that AR will play a significant role in preserving and sharing the rich heritage of human cultures.

\bibliographystyle{ACM-Reference-Format}
\bibliography{sample-base}

\end{document}